\documentclass[prd,longbibliography,twocolumn,amsmath,amssymb,nofootinbib,preprintnumbers,balancelastpage]{revtex4-1}

\usepackage[utf8]{inputenc}
\usepackage{bbm}
\usepackage{graphicx}

\DeclareUnicodeCharacter{2212}{-}

\AtBeginDocument{%
    \newwrite\bibnotes
    \def\bibnotesext{Notes.bib}
    \immediate\openout\bibnotes=\jobname\bibnotesext
    \immediate\write\bibnotes{@CONTROL{REVTEX41Control}}
    \immediate\write\bibnotes{@CONTROL{%
    apsrev41Control,author="08",editor="1",pages="1",title="0",year="1"}}
     \if@filesw
     \immediate\write\@auxout{\string\citation{apsrev41Control}}%
    \fi
}%

\usepackage{graphicx}
\usepackage{hyperref}
\hypersetup{
    pdfnewwindow=true,      
    colorlinks=true,        
    linkcolor=blue,        
    citecolor=blue,         
    filecolor=blue,         
    urlcolor=blue           
}

\newcommand{\beq}{\begin{equation}}
\newcommand{\eeq}{\end{equation}}

\newcommand{\SU}{\,{\rm SU}}

\newcommand{\U}{\,{\rm U}}

\begin{document}

\title{\Large {\bf{Electric Dipole Moments, New Forces and Dark Matter}}}
\author{Pavel Fileviez P\'erez}
\email{pxf112@case.edu}
\affiliation{Physics Department and Center for Education and Research in Cosmology and Astrophysics (CERCA), 
Case Western Reserve University, Cleveland, OH 44106, USA}
\author{Alexis D. Plascencia}
\email{alexis.plascencia@case.edu}
\affiliation{Physics Department and Center for Education and Research in Cosmology and Astrophysics (CERCA), 
Case Western Reserve University, Cleveland, OH 44106, USA}

\begin{abstract}
New sources of CP violation beyond the Standard Model are crucial to explain the baryon asymmetry in the Universe. We discuss the impact of new CP violating interactions in theories where a dark matter candidate is predicted by the cancellation of gauge anomalies. In these theories, the constraint on the dark matter relic density implies an upper bound on the new symmetry breaking scale from which all new states acquire their masses. We investigate in detail the predictions for electric dipole moments and show that if the relevant CP-violating phase is large, experiments such as the ACME collaboration will be able to fully probe the theory.
\end{abstract}

\maketitle

\section{Introduction}
The search for violation of the CP symmetry in nature represents a powerful tool to search for physics beyond the Standard Model (SM). 
The existence of new large CP-violating phases beyond the SM are needed to explain the observed matter-antimatter asymmetry in the Universe. 
Unfortunately, the baryon asymmetry cannot be explained in the context of the SM even if CP is broken in the quark sector.

An important observable that arises from the violation of the CP symmetry is the electric dipole moment (EDM) of elementary particles. 
Recently, the ACME collaboration has set an impressive new upper limit~\cite{Andreev:2018ayy}: $$\frac{|d_e|}{e}  < 1.1 \times 10^{-29} \,\, {\rm cm},$$ 
on the electron electric dipole moment. For reviews on this subject we refer the reader to Refs.~\cite{Bernreuther:1990jx,Pospelov:2005pr,Fukuyama:2012np,Chupp:2017rkp}. There is a 
large list of studies in this field. Previous studies about CP violation and the predictions for EDMs have mostly focused in the context of the Minimal Supersymmetric Standard Model (MSSM), see e.g.~\cite{Ibrahim:1998je,Abel:2001vy,Lebedev:2002ne,Chang:2002ex,Pilaftsis:2002fe,Demir:2003js,Carena:2004ha,Li:2008kz,Ellis:2008zy,Yamanaka:2012ia,McKeen:2013dma,Nakai:2016atk,Cesarotti:2018huy}, and split-SUSY~\cite{ArkaniHamed:2004yi,Chang:2005ac,Giudice:2005rz}; there are a few studies in the context of dark sectors~\cite{Fuyuto:2019vfe,Okawa:2019arp} and see Refs.~\cite{Fan:2013qn,Cirigliano:2016nyn,Abe:2017glm,Alioli:2017ces,Frigerio:2018uwx,Panico:2018hal} for other studies. The current experimental upper bounds on the EDMs already constrain new physics at the TeV scale if one has large CP-violating phases.

The predictions for EDMs in theories for physics beyond the SM depend on two main factors: 1) the new CP-violating phases and 2) the new scale defining 
the mass of the fields generating the EDMs, see Refs.~\cite{Bernreuther:1990jx,Pospelov:2005pr,Fukuyama:2012np,Chupp:2017rkp} for more details.
For example, in supersymmetric theories there can be new CP-violating phases coming from the supersymmetry 
(SUSY) breaking sector and the SUSY scale defines the overall scale of all superpartner masses. 
However, there exist different scenarios such as split-SUSY, and in general, the SUSY scale can be high unless we are restricted to be in a low-energy SUSY scenario.
Generically, the mass scale associated to the generation of the EDMs can be pushed up to very high values in many theories for physics beyond the SM and the predictions for EDMs can be far from the current experimental bounds.

In this letter, we investigate the predictions for EDMs in the context of gauge theories where a dark matter candidate is predicted by the cancellation of gauge anomalies. 
In this context one predicts the existence of new CP-violating phases and the cosmological bounds on the dark matter relic density implies an upper bound on the new 
symmetry scale in the multi-TeV. We show that one can predict large values for the electron EDM in this context if the CP-violating phases are large.

A dark matter candidate is predicted in simple gauge theories where the anomaly cancellation predicts a new electrically neutral field which is automatically stable after symmetry breaking.
The minimal theories with this prediction correspond to promoting baryon and/or lepton number to local gauge symmetries~\cite{Duerr:2013dza,Perez:2014qfa}. We discuss in detail the predictions for EDMs in the minimal theory that describes the spontaneous breaking of local baryon number at the low scale. In these theories, the existence of an upper bound on the dark matter mass implies an upper bound on the full theory since all particles acquire a mass from the same symmetry breaking scale. Consequently, the charged fermions responsible for the EDMs must live below the multi-TeV scale which leads to large values for the electron EDM that can be fully probed in the near future. Similar results can be obtained in other gauge theories with these features.

This letter is organized as follows. In Section~\ref{sec:DM}, we discuss the connection to dark matter and the upper bound on the symmetry breaking scale. In Section~\ref{sec:theory}, we discuss the general aspects of gauge extensions of the SM that give rise to a dark matter candidate from the cancellation of gauge anomalies. In Section~\ref{sec:EDMs}, we show that CP violation is present in these theories and calculate the contribution to the electric dipole moments of SM fermions from the two-loop Barr-Zee diagrams~\cite{Barr:1990vd} shown in Fig.~\ref{fig:FeynmanDiag}. We summarize our findings in Section~\ref{sec:Summary}. Further details are provided in the Appendices~\ref{sec:EDMcalc}, \ref{sec:appmasses} 
and \ref{sec:appFR}.
%
\section{EDMs and Dark Matter}
\label{sec:DM}
%
In theories beyond the SM with new gauge forces one typically needs a new sector to define an anomaly free theory. This new sector can provide new sources for CP violation and if the new particles are not very heavy then  large values for the electric dipole moments can be predicted. For example, in theories based on gauging baryon $\U(1)_B$ or lepton number $\U(1)_L$, the new sector must be light because one of the fields needed for anomaly cancellation is a cold dark matter candidate. The cosmological constraint on the dark matter relic density, $\Omega_{\text{DM}} h^2 \leq 0.12$~\cite{Aghanim:2018eyx}, implies that the dark matter candidate must be below the multi-TeV scale~\cite{FileviezPerez:2019jju, FileviezPerez:2019cyn}. In these theories, all the new fermions acquire mass from the new symmetry scale, and hence, the fields that contribute to the EDMs must also be light.

\begin{figure}[h]
\centering
\includegraphics[width=0.65\linewidth]{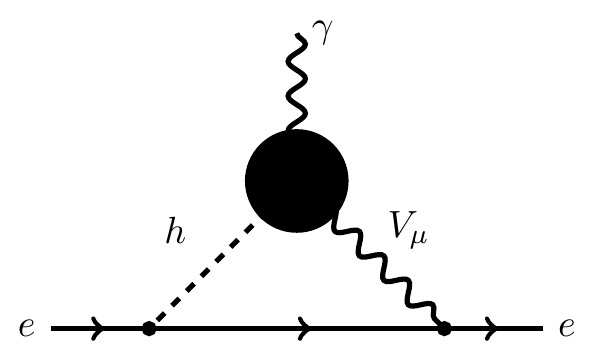}
\caption{Barr-Zee contribution to the electric dipole moment of the electron. $V_\mu$ is a generic neutral gauge boson, e.g. the photon, the $Z$ or a $Z'$. Here $h$ corresponds to the SM Higgs boson.}
\label{fig:FeynmanDiag}
\end{figure}

In Fig.~\ref{fig:FeynmanDiag} we show the leading contribution to the electron dipole moment in the theories mentioned above, where in the `black-box' (circle) one has the new fermions needed for anomaly cancellation that carry $\SU(2)_L$ and $\U(1)_Y$ quantum numbers. 
In order to investigate the correlation  between the predictions for the electron electric dipole moment and the dark matter constraints we discuss a simple model where the existence of dark matter is predicted and there exist new sources of CP violation. 

It is important to mention that in other theories, such as supersymmetric ones, there can also be a connection between the predictions for EDMs and the properties of the lightest neutralino as a dark matter candidate; however, these theories are more involved and the relevant scale cannot be predicted. Therefore, it is fair to say that large values of the EDMs are not a general prediction of the MSSM.

\section{Theoretical Framework}
\label{sec:theory}
In theories for physics beyond the SM where a new gauge symmetry is spontaneously broken there can be new sources of CP-violation. In this article, we focus on extensions of the SM where the new symmetry is not anomaly free in the SM. The simplest cases correspond to the cases where B and/or L are promoted to local gauge symmetries~\cite{Perez:2014qfa,Duerr:2013dza}. Our main results can be applied to different theories but to make the discussion clear we will show the predictions in the context of the minimal theory for local baryon number. 

An anomaly free gauge theory for baryon number can be defined by adding only 
four extra representations~\cite{Perez:2014qfa}. The extra fermion fields needed for anomaly cancellation are the following:
\begin{eqnarray}
\Psi_L \sim (\mathbf{1},\,\mathbf{2},\,1/2,\,3/2), \quad \Psi_R \sim (\mathbf{1},\,\mathbf{2},\,1/2,\,-3/2), \nonumber \\
\quad \Sigma_L \sim (\mathbf{1},\,\mathbf{3},\,0,\,-3/2), \quad {\rm and} \quad \chi_L \sim (\mathbf{1},\,\mathbf{1},\,0,\,-3/2). \nonumber
\end{eqnarray}
Notice that this theory predicts the baryon number for each multiplet to be $\pm 3/2$ and the existence of a Majorana dark matter candidate. 
In this context, the lightest electrical neutral Majorana field is our dark matter candidate.  

The Yukawa interactions in this theory for the new fermions are given by
\begin{eqnarray}
\label{eq:Yukawas}
-\mathcal{L} \  & \supset \ &  y_1  \bar{\Psi}_R H \chi_L + y_2 H^\dagger \Psi_L \chi_L  + y_3 H^\dagger \Sigma_L  \Psi_L \nonumber \\ 
& + & y_4  \bar{\Psi}_R \Sigma_L H  +  y_\Psi \bar{\Psi}_R \Psi_L S^*_B \nonumber \\
&+ & \frac{y_\chi}{\sqrt{2}} \chi_L \chi_L S_B + y_\Sigma {\rm Tr} (\Sigma_L \Sigma_L) S_B + {\rm h.c.},
\end{eqnarray}
where $H \sim (\mathbf{1}, \mathbf{2},1/2,0)$ corresponds to the Higgs doublet in the SM and the scalar $S_B \sim (\mathbf{1}, \mathbf{1},0,3)$ acquires a non-zero vacuum expectation value 
generating the masses for the new states. We define the mass parameters
\beq
\mu_\Sigma = \sqrt{2} y_\Sigma v_B, \hspace{1cm} \mu_\Psi = \frac{y_\Psi}{\sqrt{2}} v_B,
\eeq
and in Appendix~\ref{sec:appmasses} we discuss the mass matrices and their diagonalization to find the physical masses.
It is important to emphasize that local baryon number must be broken in 3 units, and hence, the proton is predicted to be stable and the symmetry breaking scale can be low.

After symmetry breaking, the local $\U(1)_B$ is broken to a ${\mathcal{Z}}_2$ symmetry which protects the dark matter candidate from decaying. After symmetry breaking, one has the symmetry: 
$$\{ \Psi_L \to - \Psi_L , \ \Psi_R \to - \Psi_R, \  \Sigma_L \to - \Sigma_L , \chi_L^0 \to - \chi_L^0 \}.$$
Therefore, if the lightest field in this sector is neutral we have a scenario consistent with cosmology and a cold dark matter candidate.
This simple theory predicts the existence of four neutral fermions $\chi_i^0$ $(i=1,2,3,4)$, two charged fermions $\chi^+_j$ $(j=1,2)$, a neutral 
Higgs boson $h_B$ and a neutral gauge boson $Z_B$. The lightest neutral fermion $\chi_1^0$ is stable and describes the cold dark matter in the Universe.
See Appendix~\ref{sec:appmasses} for a detailed discussion of the properties of the new fermionic states. 
\begin{figure}[b]
\centering
\includegraphics[width=0.93\linewidth]{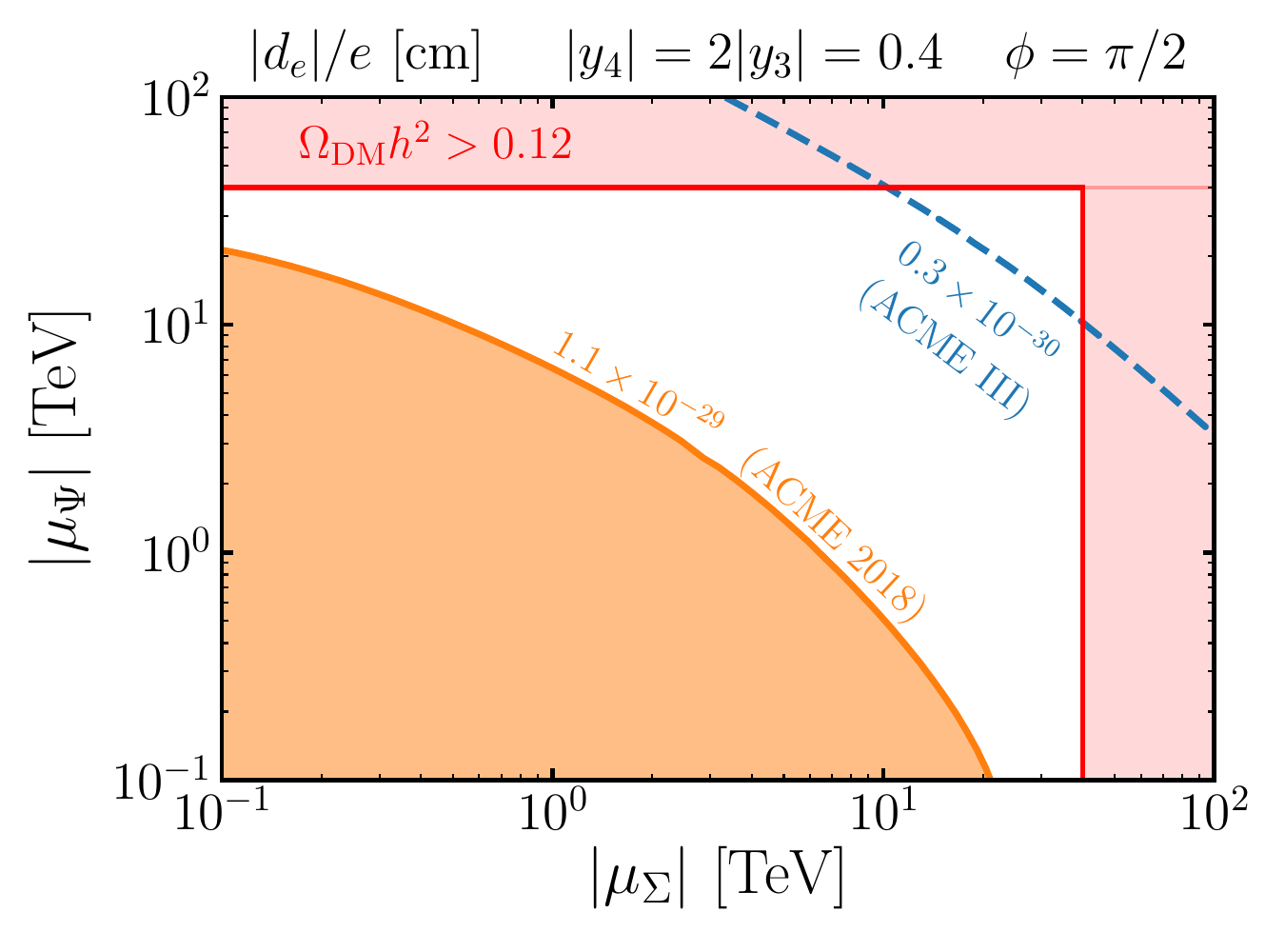}
\includegraphics[width=0.93\linewidth]{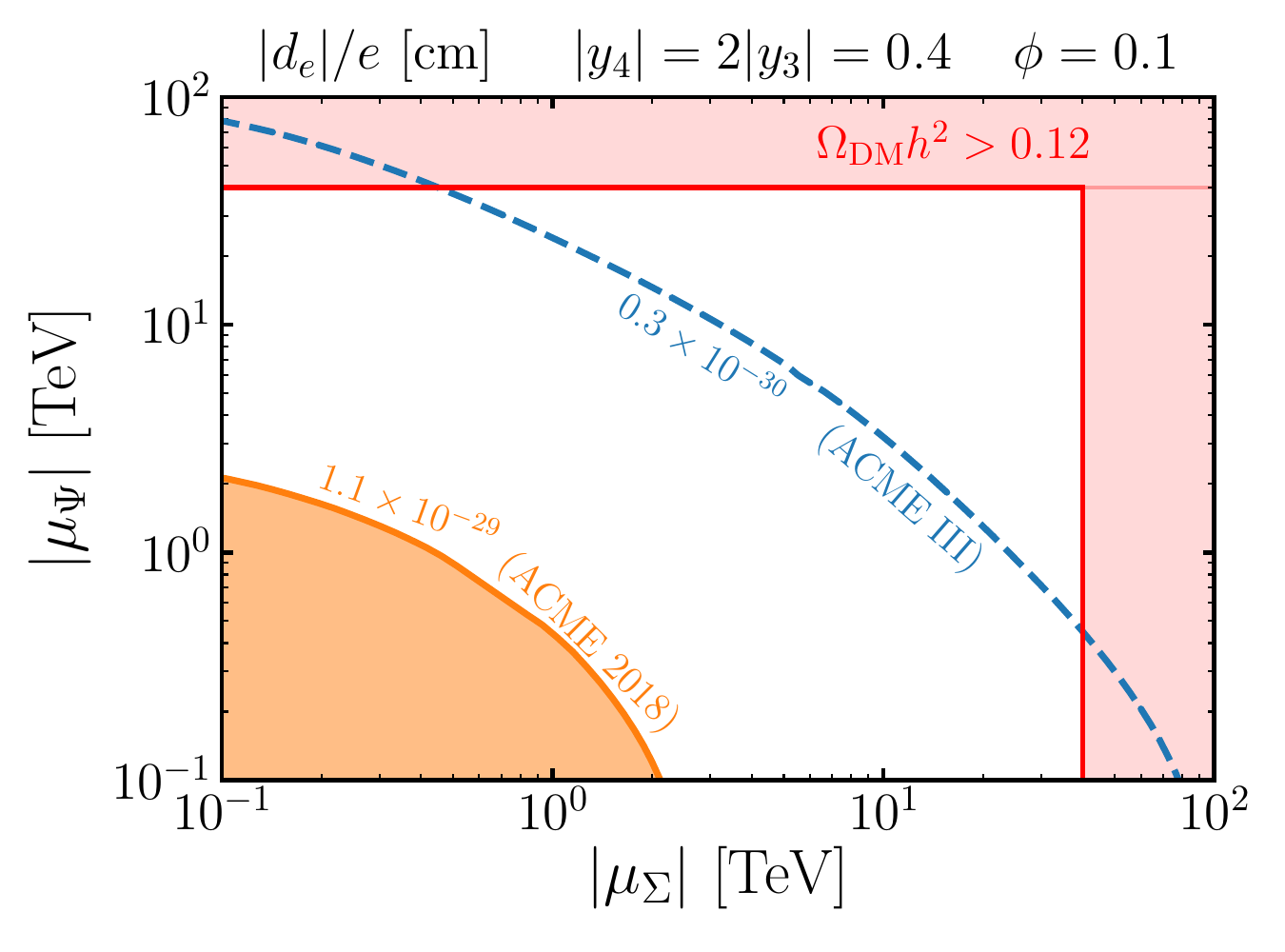}
\caption{Contour plots for the EDM of the electron, $|d_e|/e$, in units of centimeters in the $|\mu_\Psi|$ vs  $|\mu_\Psi|$ plane. The region shaded in orange is excluded by the ACME  result~\cite{Andreev:2018ayy} and the blue dashed line gives the projected sensitivity for ACME III~\cite{ACMEIII}. The region shaded in light red corresponds to overproduction of dark matter. We have taken zero scalar mixing angle. The upper (lower) panel corresponds to a CP-violating phase $\phi={\rm arg} (y_3^*y_4^* \mu_\Sigma \mu_\Psi) =\pi/2$ ($\phi=0.1$). }
\label{fig:eEDM}
\end{figure}

The new Yukawa couplings present in Eq.~\eqref{eq:Yukawas}, $y_i$ $(i=1..4)$, $y_{\Psi}$, $y_{\chi}$, and $y_\Sigma$ are in general complex and there are new sources of CP violation. It is straightforward to see that there are three CP-violating phases in this theory that cannot be rotated away. In the charged fermionic sector there is only one CP-violating phase, $\phi={\rm arg} (y_3^*y_4^* \mu_\Sigma \mu_\Psi)$, relevant for the EDM predictions. 
\section{Predictions for EDMs}
\label{sec:EDMs}
The ACME Collaboration has found an upper limit on the electron EDM at 90$\%$ confidence level~\cite{Andreev:2018ayy}
\beq
\frac{|d_e|}{e}  < 1.1 \times 10^{-29} \,\, {\rm cm}.
\eeq
The main contribution to the electron EDM in the class of theories we study is shown in Fig.~\ref{fig:FeynmanDiag} with $V_\mu=\gamma$ and it is given by
\begin{align}
\label{eq:degammah}
d_e^{\gamma h} & =  \frac{\alpha^2 \cos \theta_B \, Q_e \, m_e}{8 \pi^2 s_W \, m_h^2 \, m_W}   \sum_{i=1}^2 M_{\chi_i^\pm} \, {\rm Im} [ C_h^{ii} ] \, I_{\gamma h}^i(M_{\chi_i^\pm}), 
\end{align}
where the loop integrals $I(m)$ are given in Appendix~\ref{sec:EDMcalc} and the Feynman rules used to derive this expression are given in Appendix~\ref{sec:appFR}. There are also contributions from having $hZ$, $hZ'$, and $WW$ in the loop; however, these contributions are subleading when compared to the $\gamma h$ contribution. The contributions involving $h_B$ will be suppressed by the mixing angle and by its large mass $M_{h_B}$.  In the above equation $M_{\chi_i^\pm}$ correspond to the physical masses of the anomaly-canceling charged fermions 
and $\theta_B$ is the mixing angle between the two Higgses present in the theory.
\begin{figure}[t]
\centering
\includegraphics[width=0.97\linewidth]{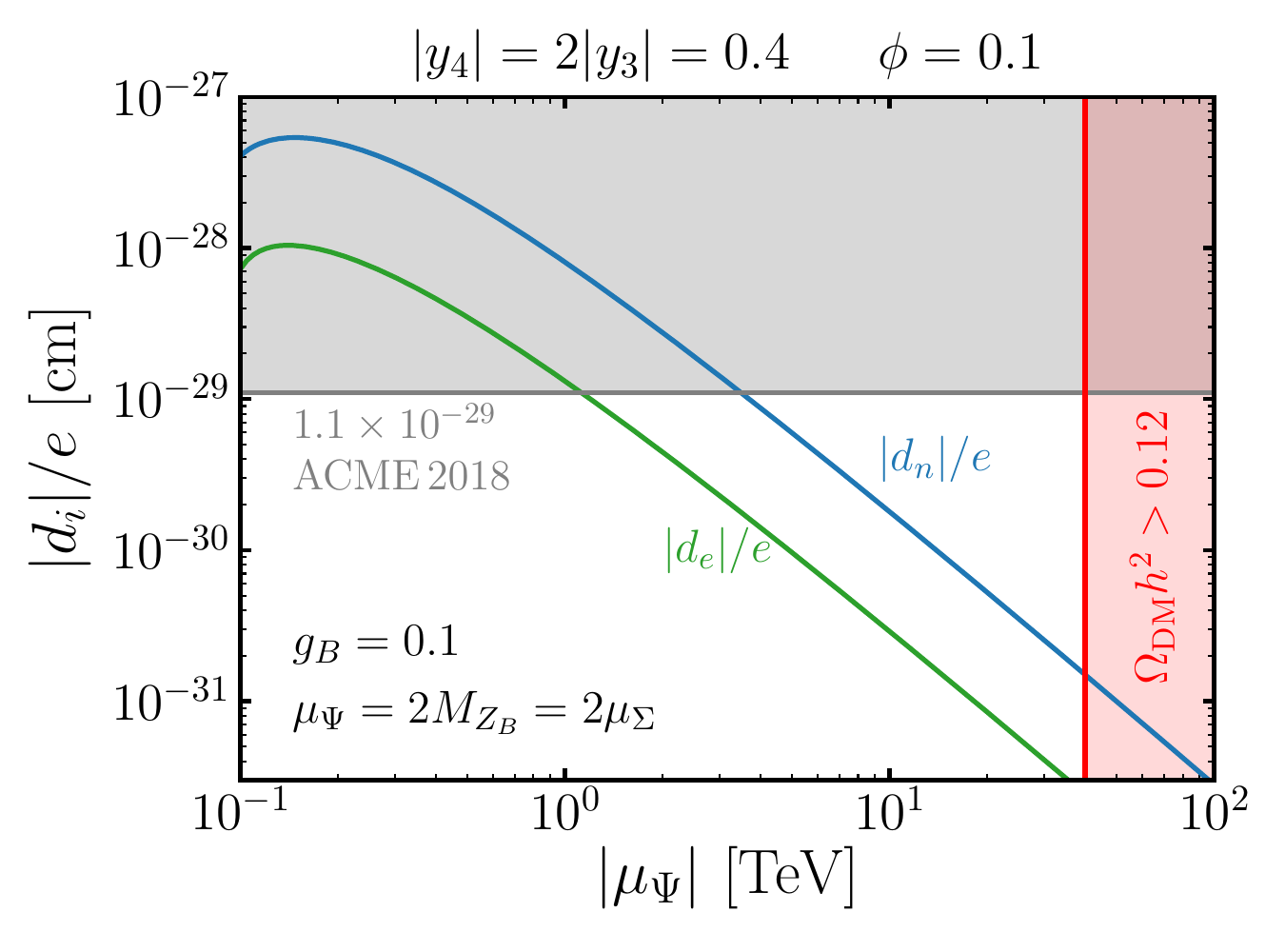}
\caption{Predictions for the EDMs as a function of the $|\mu_\Psi|$ parameter. The blue (green) line corresponds to the EDM of the neutron (electron). The region shaded in gray is the upper limit on the electron EDM obtained by the ACME Collaboration~\cite{Andreev:2018ayy}. The region shaded in light red corresponds to overproduction of dark matter. The Yukawa couplings and the mass parameters are fixed as shown in the figure. The CP-violating phase has been set to $\phi={\rm arg} (y_3^*y_4^* \mu_\Sigma \mu_\Psi) = 0.1$.}
\label{fig:EDMs}
\end{figure}

The electron EDM in Eq.~\eqref{eq:degammah} depends only on one CP-violating phase $\phi={\rm arg}(y_3^* y_4^* \mu_\Sigma \mu_\Psi)$. In Fig.~\ref{fig:eEDM}, we show the predictions of the EDM of the electron. For a maximal CP-violating phase of $\phi=\pi/2$ the ACME constraint already excludes masses for the charged fermions of up to 20 TeV, by reducing the phase to $\phi=0.1$ this exclusion goes down to 2 TeV. The region shaded in red corresponds to the upper bound on the fermion masses that will be discussed below. For a large CP-violating phase, the predictions for the electron EDM in these theories  can be fully probed in the near future~\cite{ACMEIII,Kozyryev:2017cwq}. 

In order to show the dependence on the mass paramaters, in Fig.~\ref{fig:eEDM} we have fixed the value of the Yukawas to $|y_4|=2 |y_3|=0.4$. As we discuss below, the calculation of the dark matter relic density does not depend on these parameters. Nonetheless, from the results in Appendix~\ref{sec:EDMcalc} it can be seen that in the limit in which these Yukawa couplings vanish then the EDMs will also vanish.

The EDM of the neutron is given by \cite{Pospelov:2005pr}
\beq
d_n = (1.4 \pm 0.6) \, (d_d - 0.25 d_u),
\eeq
where $d_{d,u}=d_{d,u}^{\gamma h}+d_{d,u}^{hZ}+d_{d,u}^{hZ'}$ correspond to the EDMs of the down and up quarks and include the different contributions from $\gamma h$, $h Z$ and $h Z'$, the explicit expressions are given in Appendix~\ref{sec:EDMcalc}. As pointed out in Ref.~\cite{Giudice:2005rz}, the $hZ$ contribution can be comparable to the $\gamma h$ contribution for the neutron EDM. In the above equation we do not include the contribution proportional to $\bar{\theta}_{\rm QCD}$. In Fig.~\ref{fig:EDMs} we show our results, the blue line corresponds to the predictions for the neutron EDM while the green line is the one for the electron EDM. The current experimental bound is $|d_n| <1.8\times 10^{-26} \,\, e\,\, {\rm cm}$~\cite{Abel:2020gbr} which is much weaker than the one for the electron EDM and does not give rise to independent constraints.

\begin{figure}[t]
\centering
\includegraphics[width=0.93\linewidth]{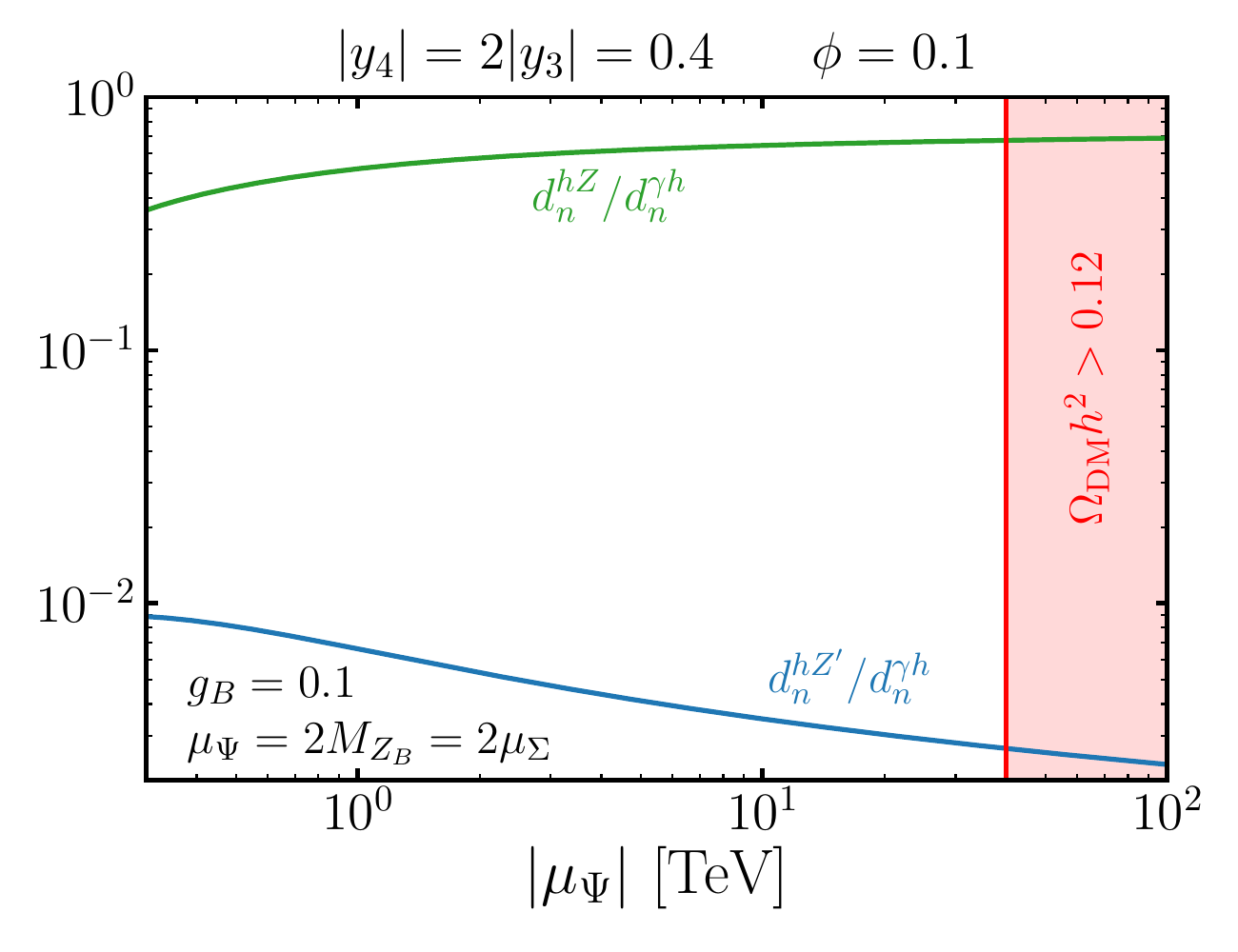}
\caption{Ratios of the different contributions to the EDM of the neutron as a function of $|\mu_\Psi|$. The other mass parameters are fixed in relation to $|\mu_\Psi|$ as shown in the figure. The region shaded in light red corresponds to overproduction of dark matter.  We have taken zero scalar mixing angle. We have fixed the CP-violating phase to $\phi={\rm arg} (y_3^*y_4^* \mu_\Sigma \mu_\Psi) = 0.1$.}
\label{fig:ratios}
\end{figure}

In the theories we have discussed, there is new gauge boson $Z'$ interacting with the SM fermions that gives rise to a new contribution to the EDM; namely, by having $V_\mu\!=\! Z'$ in Fig.~\ref{fig:FeynmanDiag}. In theories for local lepton number, $\U(1)_L$,  this gives a new contribution to $d_e$; however, due to the strong collider~\cite{Alioli:2017nzr,Aaboud:2017buh} and cosmological~\cite{FileviezPerez:2019cyn} constraints on a $Z'$ coupled to leptons this contribution is highly suppressed. In the $\U(1)_B$ scenario the gauge boson can live at a low scale with a large gauge coupling, and hence, the contribution to $d_n$ can be significant. See Ref.~\cite{FileviezPerez:2020mtk} for a recent discussion of the collider bounds for the gauge boson associated to baryon number. In Fig.~\ref{fig:ratios} we present the ratios $d_h^{hZ}/d_n^{\gamma h}$ and $d_h^{hZ'}/d_n^{\gamma h}$ to show how the contribution from these channels compare to the $\gamma h$ contribution.

\begin{figure}[b]
\centering
\includegraphics[width=0.93\linewidth]{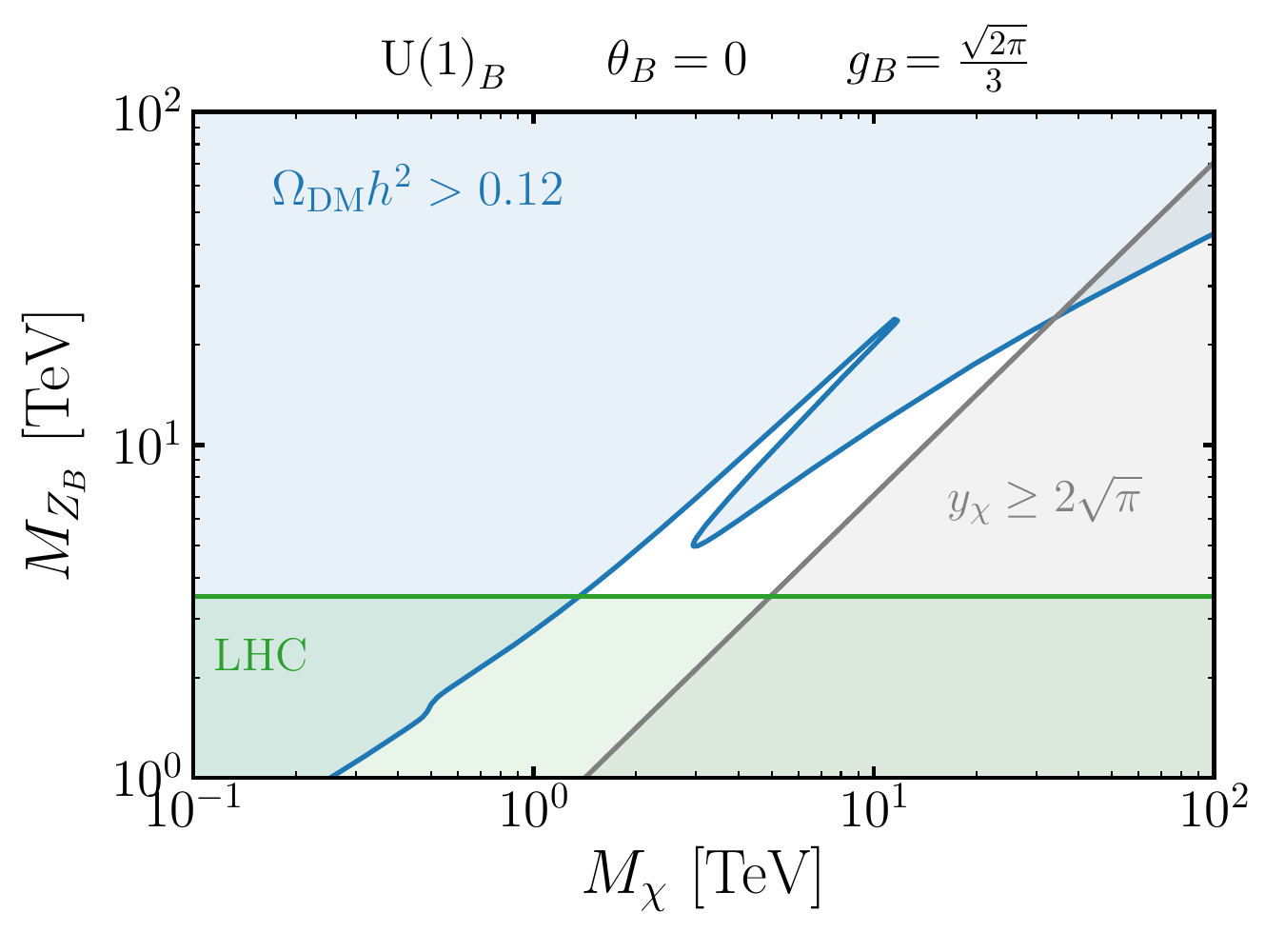}
\caption{Parameter space allowed by the relic density constraint, LHC bounds and perturbative bounds for the maximal gauge coupling $g_B=\sqrt{2 \pi}/3$. We take $M_{h_B}=500$ GeV and no mixing angle. The region shaded in blue is excluded by the relic density constraint $ \Omega_{\rm{DM}} h^2 \leq 0.12$ and the region in gray is excluded by the perturbative bound on the Yukawa coupling $y_\chi$. 
The region shaded in green is excluded by the LHC bounds on the leptophobic gauge boson mass.}
\label{fig:DMU1B}
\end{figure}

As mentioned previously, the lightest neutral state is a good dark matter candidate. In Fig.~\ref{fig:DMU1B} we present the relic density and the experimental constraints in the $\U(1)_B$ scenario taking $g_B\leq\sqrt{2\pi}/3$, the region shaded in blue overproduces the dark matter relic density and the gray region is excluded by the perturbativity of the Yukawa coupling  $y_\chi$. In this theory the Majorana dark matter candidate can have the following annihilation channels:
$$\chi \chi \to \bar{q}q, \, Z_B Z_B, \, Z_B h_i, \, h_i h_j, \, WW,  \, ZZ.$$
Here $h_i=h,h_B$, where $h_B$ is the second physical Higgs present in the theory.
The peak in Fig.~\ref{fig:DMU1B} corresponds to the resonance $\chi \chi \to Z_B^* \to \bar{q} q$ and away from this resonance the dominant annihilation is $\chi  \chi  \to h_B Z_B$. Consequently, the dark matter relic density is most sensitive to the following parameters: the dark matter Yukawa coupling $y_\chi$, the gauge coupling $g_B$ and the masses $M_{Z_B}$ and $M_{h_B}$. Regarding the direct detection of dark matter, due to its Majorana nature the channel mediated by the $Z_B$ is velocity-suppressed. Nonetheless, large scalar mixing angles can lead to detectable cross-sections in the near future. 
 For a detailed study of the dark matter constraints see Ref.~\cite{FileviezPerez:2019cyn}. 
 
An upper bound on the dark matter mass can be found by the requirement of not overclosing the Universe. In Refs.~\cite{FileviezPerez:2019jju, FileviezPerez:2019cyn}, by taking the maximal gauge coupling allowed by perturbativity $g'\leq\sqrt{2\pi}/3$ we found that $M_{Z'}\leq 28$ (21) TeV and $M_{\rm DM} \leq 34$ (34) TeV for the $\U(1)_B$ ($\U(1)_L$) theory. Although, due to the positive renormalization group equation for the gauge coupling $g_B$, a theory with such a large $g_B$ is unexpected since the theory will quickly run into a Landau pole; thus,  the theory is expected to be at a lower scale. 

All new fermions in the theory acquire their masses from the same $\U(1)'$ symmetry breaking scale; namely, in the limit of small fermionic mixing we have that $M_{\chi_1^\pm} \approx \sqrt{2} y_\Sigma v' = \sqrt{2}y_\Sigma M_{Z'}/3g'$ and $M_{\chi_2^\pm} \approx y_\Psi v'/\sqrt{2}=y_\Psi M_{Z'}/3\sqrt{2}g'$, then the upper bound on $M_{Z'}$ translates as an upper bound on the fermion masses. Taking the largest value of the Yukawa coupling allowed by  perturbativity ($y_\Sigma\leq \sqrt{2 \pi}$ and $y_\Psi\leq 2\sqrt{2\pi}$) we find that
\beq
M_{\chi_i^\pm} \leq 40 \,\, {\rm TeV},
\eeq
for the $\U(1)_B$ theory and shown by the red shaded region in Figs.~\ref{fig:eEDM}, \ref{fig:EDMs} and \ref{fig:ratios}. For $\U(1)_L$ this upper bound corresponds to 53 TeV.
Therefore, since there is an upper bound, these theories predict large values for the EDM of the electron as our numerical results show.

These results can have important implications for the mechanisms to explain the baryon asymmetry in the Universe. For a recent proposal of a mechanism of baryogenesis in theories with gauged baryon or lepton number see Refs.~\cite{Carena:2018cjh,Carena:2019xrr}. A detailed study of the implications of our results for the generation of the matter-antimatter asymmetry is beyond the scope of this letter.
%

%
\section{Summary}
\label{sec:Summary}
Experiments such as the ACME collaboration searching for electric dipole moments can shed light on the origin of CP violation in nature, that is crucial to explain the baryon asymmetry in the Universe.
In this letter, we have shown that there is a class of gauge theories which predict new CP-violating interactions once we try to define an anomaly free theory.
These theories predict the existence of a new sector needed for anomaly cancellation and since they contain  a thermal dark matter candidate, all states have to be below the multi-TeV scale.

We investigated in detail one of these theories, the gauge theory for baryon number, where the symmetry breaking scale should be below $M_{Z_B} \lesssim 30$ TeV.
In this theory, the dark matter candidate is a Majorana field and the other charged fields must be below the multi-TeV scale once we impose the relic density constraints. 
Therefore, this theory predicts large values for the electron EDM that can be tested in the near future. 
The presence of a new gauge boson coupled to SM fermions gives rise to a new contribution to the EDM of SM fermions that we have computed. However, since the mass of this gauge boson is related to the mass of the charged fermions that live at the TeV scale, this contribution turns out to be small. 

Our results demonstrate that for large values of the relevant CP-violating phase, experiments that search for the EDM of the electron such as ACME III can probe the predictions of these theories. These are striking results which tell us that there is hope to measure a non-zero electron EDM if these theories are realized in nature.
\vspace{0.5cm}
\\
{\textit{Acknowledgments:}}
{\small{The work of P.F.P. has been supported by the U.S. Department of Energy, Office
of Science, Office of High Energy Physics, under Award Number de-sc0020443.}}

\begin{widetext}
\appendix
\section{EDMs contributions}
\label{sec:EDMcalc}
\begin{figure}[bp]
\centering
\includegraphics[width=0.35\linewidth]{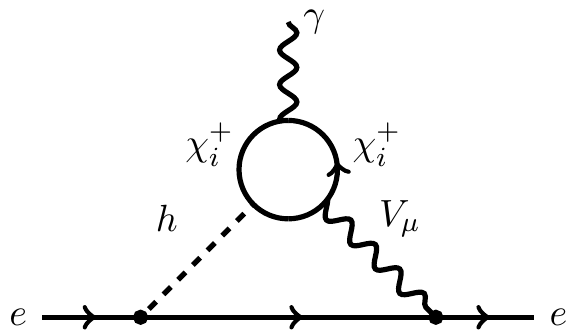}
\caption{Barr-Zee contribution to the electric dipole moment of the electron. $V_\mu$ is a generic neutral gauge boson, e.g. the photon, the $Z$ or a $Z'$. Here $\chi_i^+$ correspond to the anomaly-canceling fermions.}
\label{fig:DiagEDM}
\end{figure}
In computing the two-loops shown in Fig.~\ref{fig:DiagEDM} we follow the approach presented in Ref.~\cite{Nakai:2016atk}. 
The different contributions to the EDM of a SM fermion $f$ are given by
\begin{align}
d_f^{\gamma h} & =  \frac{\alpha^2 \cos \theta_B Q_f }{4 \pi^2 s_W } \frac{m_f}{m_h^2 \, m_W} \sum_{i=1}^2 M_{\chi_i^\pm} {\rm Im}[C_h^{ii}] \, I_{\gamma h}^i(M_{\chi_i^\pm}) ,
\\[1.5ex]
d_f^{hZ} & =  \frac{e \alpha \cos \theta_B}{32 \pi^2 c_W s_W^2} \left( \frac{1}{2}T_3^f - s_W^2 Q_f \right) \frac{m_f}{m_h^2 m_W} \sum_{i,j=1}^2  I_{hZ}^{ij}(M_{\chi_i^\pm}, M_{\chi_j^\pm}), \\[1.5ex]
d_f^{hZ'} & = \frac{ g_B
 \alpha \cos \theta_B}{96 \pi^2 s_W} \frac{m_f}{m_h^2 m_W} \sum_{i,j=1}^2  I_{hZ'}^{ij}(M_{\chi_i^\pm}, M_{\chi_j^\pm}),
\end{align}
where $M_{\chi_i^\pm}$ correspond to the physical masses of the anomaly-canceling fermions and the loop integrals $I(m)$ are given by
\begin{align}
I_{\gamma h}^i(M_{\chi_i^\pm}) = & \int_0^1  \frac{dx}{x} j\left(0, \frac{M^2_{\chi_i^\pm}}{m_h^2} \frac{1}{x(1-x)} \right), \nonumber \\[1.5ex]
I_{hZ}^{ij}(M_{\chi_i^\pm}, M_{\chi_j^\pm}) = & \int_0^1 dx \frac{1}{x(1-x)}  j\left(\frac{m_Z^2}{m_h^2
}, \frac{\Delta_{ij}(x)}{m_h^2} \right) \nonumber \\[1ex]
&  {\rm Re} \left[(i g_S^{ji} \, (g_A^{ij})^* - g_P^{ji} \, (g_V^{ij})^*) M_{\chi^\pm_i} x(1-x)   -  (i g_S^{ji} \, (g_A^{ij})^* + g_P^{ji} \, (g_V^{ij})^*) M_{\chi^\pm_j} (1-x)^2 \right], \nonumber \\[1.5ex]
I_{hZ'}^{ij}(M_{\chi_i^\pm}, M_{\chi_j^\pm}) = & \int_0^1 dx \frac{1}{x(1-x)}  j\left(\frac{M_{Z'}^2}{m_h^2
}, \frac{\Delta_{ij}(x)}{m_h^2} \right) \nonumber \\[1ex]
&  {\rm Re} \left[(i g_S^{ji} \, (\kappa_A^{ij})^* - g_P^{ji} \, (\kappa_V^{ij})^*) M_{\chi^\pm_i} x(1-x)   -  (i g_S^{ji} \, (\kappa_A^{ij})^* + g_P^{ji} \, (\kappa_V^{ij})^*) M_{\chi^\pm_j} (1-x)^2 \right], \nonumber 
\end{align}
with the parameters
\begin{align}
g_S^{ij}  =  - {\rm Re} [ C_h^{ij} ],
& \hspace{1cm}
g_P^{ij}  = \,\, {\rm Im} [ C_h^{ij} ],
\\[1.5ex]
g_V^{ij}  =  - \frac{g \sin \theta_W}{2} \left( C_L^{ij} + C_R^{ij} \right),
& \hspace{1cm}
g_A^{ij} =  \,\, \frac{g \sin \theta_W}{2} \left( C_L^{ij} - C_R^{ij}  \right),
\\[1.5ex]
\kappa_V^{ij}  = \frac{3}{4} g_B \left( O_L^{ij} + O_R^{ij} \right),
& \hspace{1cm}
\kappa_A^{ij} = \frac{3}{4} g_B \left( O_R^{ij} - O_L^{ij}  \right),
\end{align}
where the matrices $C_h$, $C_L$, $C_R$, $O_L$ and $O_R$ are given in Appendix~\ref{sec:appFR}.  
The functions in the integrands correspond to 
\begin{align}
j(y,z)= & \frac{1}{y-z} \left( \frac{y\log y}{y-1} - \frac{z\log z}{z-1}  \right), \\[1.5ex]
\Delta_{ij}(x) = &  \frac{x M_{\chi_i^\pm}^2 + (1-x) M_{\chi_j^\pm}^2}{x(1-x)}.
\end{align}
%
\section{Fermionic States}
\label{sec:appmasses}
%
\begin{itemize}

\item Neutral States: The mass matrix for the neutral states in the basis $( \chi^0_L, \Sigma^0_L, \Psi_{1L}^0, \Psi_{2L}^0 )$ is given by,
\beq
{\mathcal{M}}_0 = \begin{pmatrix}
 y_\chi v_B &  0 &  \displaystyle  \frac{y_2 v}{\sqrt{2}}   &  \displaystyle \frac{ y_1 v}{\sqrt{2}} \\[2ex]  
0  & \sqrt{2} y_\Sigma v_B  & \displaystyle - \frac{y_3 v}{2}  & \displaystyle - \frac{y_4 v}{2} \\[2ex]
\displaystyle \frac{y_2 v}{\sqrt{2}}  & \displaystyle - \frac{y_3 v}{2} & 0 & \displaystyle \frac{y_{\Psi}}{\sqrt{2}} v_B   \\[2ex]
\displaystyle \frac{ y_1 v}{\sqrt{2}} & \displaystyle - \frac{y_4 v}{2}  & \displaystyle \frac{y_{\Psi}}{\sqrt{2}} v_B & 0 
 \end{pmatrix},
\eeq
where $\Psi_{2L}^0=(\Psi_{2R}^0)^C$. The mixing matrix $N$ relates these fields to the physical mass eigenstates $\chi_i^0$ as follows
\beq
\begin{pmatrix}
\chi^0\\[1ex]
\Sigma^0\\[1ex]
\Psi_1^0\\[1ex]
\Psi_2^0
\end{pmatrix}
=
N
\begin{pmatrix}
\chi_1^0\\[1ex]
\chi_2^0\\[1ex]
\chi_3^0\\[1ex]
\chi_4^0
\end{pmatrix},
\eeq
and diagonalizes the mass matrix given above as
\beq
{\mathcal{M}}_0^{\rm diag} = N^T \ {\mathcal{M}}_0 \ N.
\eeq
Notice that the neutral sector of this theory is very similar to the neutralino sector in the MSSM.
\item Charged States: The mass matrix for the new charged fermions in the basis $\chi_L^+=(\Sigma_L^+ \  \Psi_{1L}^+ )$ and $\chi_R^+=(\Sigma_R^+ \  \Psi_{2R}^+)$ is given by
\beq
-\mathcal{L}\supset
 \begin{pmatrix}
\overline{\Sigma_R^+}  \hspace{0.3cm} &   \overline{\Psi_{2R}^+}
 \end{pmatrix} 
 \mathcal{M}_C
  \begin{pmatrix}
\Sigma_L^+  \\[2ex]  \Psi_{1L}^+
 \end{pmatrix}  + 
\text{h.c.} ,
\eeq
where
\beq
\mathcal{M}_C = \begin{pmatrix}
  \sqrt{2} y_\Sigma v_B \hspace{0.2cm} & \displaystyle \frac{y_3 v}{\sqrt{2}} \\[2ex]  \displaystyle \frac{y_4 v}{\sqrt{2}} \hspace{0.2cm} &  \displaystyle \frac{y_\Psi v_B}{\sqrt{2}}  \end{pmatrix}.
\eeq
In the above equations $\Sigma_R^+=(\Sigma_L^-)^C$.
To obtain the physical fields $\chi_i^\pm$ the mass matrix needs to be diagonalized.
The relation between the fields in the Lagrangian and the physical fields is given by the $V_L$ and $V_R$ mixing matrices
\beq
\begin{pmatrix}
\Sigma_L^+  \\[2ex]   \Psi_{1L}^+
 \end{pmatrix} = V_{L} 
 \begin{pmatrix}
  \chi_{1L}^+ \\[2ex] \chi_{2L}^+   
 \end{pmatrix} ,
 \hspace{2cm}
 \begin{pmatrix}
\Sigma_R^+  \\[2ex]   \Psi_{2R}^+ 
 \end{pmatrix} = V_{R} 
 \begin{pmatrix}
  \chi_{1R}^+ \\[2ex] \chi_{2R}^+   
 \end{pmatrix} .
\eeq
The unitary matrices $V_L$ and $V_R$ diagonalize the mass matrix as follows
\beq
V_R^\dagger \mathcal{M}_C V_L = \mathcal{M}_C^{\rm diag},
\eeq
and the following relations can be used to find $V_L$ and $V_R$
\beq
|\mathcal{M}_C^{\rm diag}|^2=V_L^\dagger \mathcal{M}_C^\dagger \mathcal{M}_C V_L = V_R^\dagger \mathcal{M}_C \mathcal{M}_C^\dagger V_R.
\eeq

\end{itemize}
\section{Feynman Rules}
\label{sec:appFR}
The Feynman rules needed for the calculation of the EDMs correspond to:

\begin{align}
\includegraphics[scale=0.65]{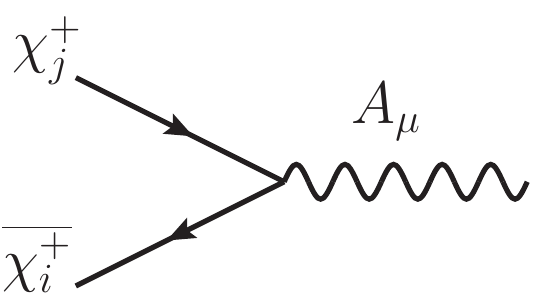} \hspace{3cm} \includegraphics[scale=0.65]{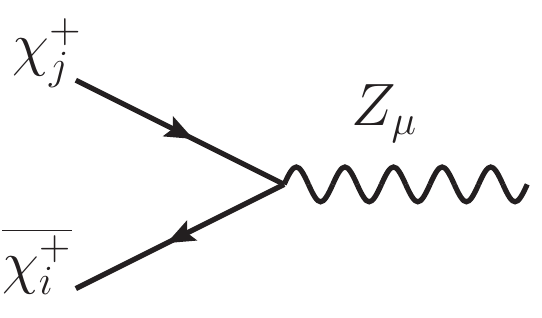} \nonumber \\
 -ie\delta_{ij}\gamma^\mu \hspace{3cm} -i g \sin \theta_W \left( C_L^{ij} P_L +  C_R^{ij} P_R \right) \gamma^\mu \nonumber
\end{align}
\\ 
\begin{align}
\includegraphics[scale=0.65]{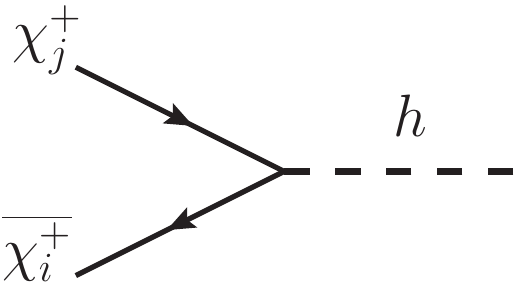} \hspace{3cm} \includegraphics[scale=0.65]{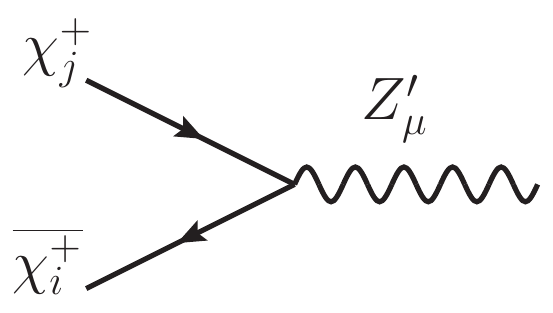} \nonumber \\
- i \left[ C^{ij}_h P_L + (C^{ij}_h)^* P_R \right] \hspace{3cm} \frac{3}{2} i g_B \left( O^{ij}_L P_L + O^{ij}_R P_R \right) \gamma^\mu \nonumber
\end{align}
\\ \newline where
\begin{align}
C_L^{ij} =& \frac{1}{\tan 2 \theta_W} (V_L^{2i})^* V_L^{2 j} + \frac{1}{\tan \theta_W} (V_L^{1i})^* V_L^{1j},
\\[1.5ex] 
C_R^{ij} =& \frac{1}{\tan 2 \theta_W} (V_R^{2i})^* V_R^{2 j} + \frac{1}{\tan \theta_W} (V_R^{1i})^* V_R^{1j},
\\[1.5ex]
C^{ij}_h =& \frac{1}{\sqrt{2}} \cos \theta_B \left[ y_3 (V_R^{1i})^* V_L^{2j} + y_4 (V_R^{2i})^* V_L^{1j} \right] \nonumber \\
\phantom{=}& +\frac{1}{\sqrt{2}} \sin \theta_B \left[ y_\Psi (V_R^{2i})^* V_L^{2j} + 2 y_\Sigma (V_R^{1i})^* V_L^{1j} \right],
\\[1.5ex]
O^{ij}_L = & (V_L^{1i})^* V_L^{1j} - (V_L^{2i})^* V_L^{2j},
\\[1.5ex]
O^{ij}_R = & (V_R^{2i})^* V_R^{2j} - (V_R^{1i})^* V_R^{1j}.
\end{align}

\end{widetext}

\bibliography{EDM-U1B}

\end{document}